\documentstyle[aps,multicol,prl,tabularx,graphics,rotate,epsf]{revtex}
\begin{document}
\title{Tracer diffusivity in a time or space dependent temperature field}
\author{Ramin Golestanian$^{1,2,3}$ and Armand Ajdari$^{2}$}
\address{$^{1}$Institute for Advanced Studies in Basic Sciences, Zanjan,
45195-159, Iran\\
$^{2}$Laboratoire de Physico-Chimie Th\'eorique, UMR CNRS 7083, E.S.P.C.I., 75231 Paris Cedex
05, France\\
$^{3}$Institute for Studies in Theoretical Physics and Mathematics,
P.O. Box 19395-5531, Tehran, Iran}
\date{to appear in Europhys. Lett.}
\maketitle
\begin{abstract}
The conventional assumption that the self-diffusion coefficient of
a small tracer can be obtained by a local and instantaneous
application of Einstein's relation in a temperature field with
spatial and temporal heterogeneity is revisited. It is shown that
hydrodynamic fluctuations contribute to the self-diffusion tensor
in a {\em universal} way, i.e. independent of the size and shape
of the tracer. The hydrodynamic contribution is {\em
anisotropic}---it reflects knowledge of the global anisotropy in
the temperature profile, leading to anisotropic self-diffusion
tensor for a spherical tracer. It is also {\em retarded}---it
creates memory effects during the diffusion process due to
hydrodynamic interactions.

\end{abstract}
\begin{multicols}{2}

{\em Introduction -} Motion of particles in thermal gradients is a
relatively old subject, of both fundamental and applied interest.
The net directed motion in a steady gradient (usually referred to
as the Soret effect or thermophoresis) has been observed and
quantified for various systems (polymer molecules, solid particles
or bubbles in a fluid) \cite{Mac} but often through indirect
measurements. However, the microscopic mechanisms responsible for
this motion are far from clear in many cases (see e.g. Ref.
\cite{Schimpf}). The resulting collective motion is usually
rationalized through Onsager-like relations, with kinetic
coefficients that are often empirically deduced from indirect
measurements. The corresponding experiments are often difficult to
set up due, in particular, to the buoyancy driven convective
effects induced by thermal gradients.

In principle, the  development of microfluidics (see
\cite{Mic1,Mic2} and references therein) should allow to create
fluidic environment less prone to such convective instabilities,
where temperature gradients can be induced and controlled, and
that should allow direct measurement of particle motion. Various
fluorescence techniques have indeed been developed that allow a
rather precise detection of temperature \cite{Ross}, and a laser
in the IR range can offer a versatile way of inducing these
gradients.

Although the directed motion of particles (Soret effect) is the
most obvious target of such a study, its amplitude often scales
with a high power of the size $a$ of the particle (typically
$a^3$), so that for small tracers an equally interesting question
is that of the diffusivity of the particle. In this paper, we take
a first step in this direction on the theory side, and use the
simplest model possible (an incompressible liquid in the Stokes
limit with thermal fluctuations of the velocity field, where the
temperature profile is externally imposed and not modified by
velocity fluctuations, and where viscosity and density are assumed
to be temperature independent), to study the effect of spatial and
temporal variations of the temperature on the diffusivity of a tracer
\cite{Pro84,Schmitz}.\\

{\em Main results -} More precisely, we re-examine the assumption
of local and instantaneous dependence of diffusion coefficient on
temperature, by taking into account the effect of hydrodynamic
fluctuations at low Reynolds number. We have studied the diffusion
of an arbitrarily shaped particle (tracer) with mobility tensor
$\mu_{ij}$ in a fluid with mass density $\rho$ and viscosity
$\eta$, which undergoes hydrodynamic fluctuations that are driven
by a given time and space dependent temperature profile $T({\bf
r},t)$. (For example, $\mu_{ij}=\delta_{ij}/(6 \pi \eta a)$ for a
spherical particle of radius $a$, while it can be a more
complicated anisotropic tensor for an arbitrarily shaped
particle.) We find that the Einstein formula for the
self-diffusion tensor takes on the form
\begin{equation}
D_{ij}({\bf r},t)=\mu_{ij} k_{\rm B} T({\bf r},t)+\delta D_{ij}({\bf r},t),     \label{Stokes}
\end{equation}
with a (usually small) correction term that is {\em universal},
i.e. independent of the shape of the tracer particle
\cite{regular}. This term is caused by an interplay between the
long-ranged hydrodynamic interactions and the spatial or temporal
heterogeneity of the temperature field. For a steady space
dependent temperature profile $T({\bf r})$, we find
\begin{eqnarray}
\delta D_{ij}({\bf r})={1 \over 32 \pi^2 \eta} \int d^3 {\bf r}' &&\; {k_{\rm B} T({\bf r}')\over
|{\bf r}-{\bf r}'|^4} \nonumber \\
\times&&\left[\delta_{ij}+2 {(r_i-r_i')(r_j-r_j')\over |{\bf r}-{\bf r}'|^2}\right],\label{dD(r)}
\end{eqnarray}
which implies that a global anisotropy in the temperature profile
will be reflected in the self-diffusion of a tracer. For the case
of a temporally varying temperature $T(t)$, on the other hand, we
obtain
\begin{equation}
\delta D_{ij}(t)={\rho^{1/2} \; \delta_{ij} \over 4 (2\pi \eta)^{3/2}} \int_{-\infty}^t d t'
\; {k_{\rm B} T(t')\over (t-t')^{3/2}},\label{dD(t)}
\end{equation}
which manifests a memory effect corresponding to retardation in hydrodynamic flows. \\

{\em Model -}
We consider an incompressible fluid at low Reynolds number that is subject
to a thermal random stirring force (density)
${\bf f}({\bf r},t)$. The stochastic dynamics of the velocity field ${\bf v}({\bf r},t)$,
which is subject to the
incompressibility constraint $\nabla \cdot {\bf v}=0$,
is then governed by the linearized Navier-Stokes equation
\begin{equation}
\rho \partial_t {\bf v}=- \nabla p+\eta \nabla^2 {\bf v}+{\bf f},       \label{NSequ}
\end{equation}
in which $p$ is the pressure field, and the density $\rho$ and viscosity $\eta$
are taken to be temperature independent.
The noise in the above equation is assumed to be Gaussian, with
$\langle f_i({\bf r},t)\rangle=0$ and
\begin{equation}
\langle f_i({\bf r},t) f_j({\bf r}',t')\rangle=2 \eta k_{\rm B} T({\bf r},t) \delta_{ij} (-\nabla^2)
\delta^3({\bf r}-{\bf r}') \delta(t-t'),        \label{noise}
\end{equation}
where the noise correlator is chosen so as to ensure local and
instantaneous thermal equilibrium for a sufficiently slowly
varying temperature field \cite{LanLif}. Note that the structure
of the noise term should be such that it only induces
vorticity-free velocity fluctuations, so that the system is not
driven far from equilibrium by the noise \cite{Procaccia}. The
velocity field correlation function can then be calculated in a
straightforward way. We obtain
\begin{eqnarray}
\langle v_i({\bf r},t) v_j({\bf r}',t')\rangle=
2 \eta \int d t_1  && d^3 {\bf r}_1  k_{\rm B} T({\bf r}_1,t_1) \nonumber \\
 \times \nabla G_{ki}({\bf r}-{\bf r}_1,t-t_1) &\cdot&
\nabla ' G_{kj}({\bf r}'-{\bf r}_1,t'-t_1),      \label{vivj}
\end{eqnarray}
where $G_{ij}$ is the Green's function for Eq.(\ref{NSequ}) above, i.e. in Fourier space
${\tilde G}_{ij}({\bf q},\omega)=(\delta_{ij}-{\hat q}_i {\hat q}_j)/(-i \rho \omega+\eta q^2)$.

Using the result for the velocity correlation function, we can extract the diffusion coefficient of
a tracer by assuming that the velocity fluctuations of an immersed particle would follow that of the
fluid at the particle's location by way of no-slip boundary condition. This assumption is reasonable
as long as we are probing the long time behavior of the motion. We thus define the self-diffusion
tensor via
\begin{equation}
D_{ij}({\bf r},t) \equiv \int_t^{\infty} \! dt' \,\langle v_i({\bf r},t) v_j({\bf r},t')\rangle
, \label{Dijdef}
\end{equation}
where it should be clear that this Eulerian velocity
auto-correlation function decays rapidly.

Examining the expression in Eq.(\ref{vivj}) then shows that the
diffusion tensor, as defined above, is decomposed as in
Eq.(\ref{Stokes}), in which $\mu_{ij}=\int {d^3 {\bf k} \over (2
\pi)^3} (\delta_{ij}-{\hat k}_i {\hat k}_j)/(\eta k^2)$ involves a
diverging integral that is sensitive to the details at short
length scales, whereas $\delta D_{ij}$ is a long-ranged
convolution of a hydrodynamic response function and the
temperature field. Note that $\delta D_{ij}=0$ for a constant
temperature profile. One can easily see that in fact $\mu_{ij}$
corresponds to the mobility tensor of the particle, which we would
obtain if we treated the proper hydrodynamic problem of flow past
a body, as the form of the integral expression suggests if we use
a cutoff of the order of the size of the particle \cite{ramin}. In
short, we can say that the diffusivity tensor $D_{ij}({\bf r},t)$
obtained from Eqs.(\ref{vivj}) and (\ref{Dijdef}) consists of two
contributions: (1) a term that is sensitive to short length-scale
features and thus to the geometry and size of the tracer, which
depends only locally on temperature, and (2) a term that is
sensitive to the non-local features in the temperature profile,
and is thus independent of the size and shape of the tracer as
long as these features have characteristic length-scales that are
much larger than the size of the tracer. While the closed form
expression for $\delta D_{ij}({\bf r},t)$ can be obtained for an
arbitrary temperature profile, we choose to report here separately
the results for space dependent temperature as in
Eq.(\ref{dD(r)}), and for time dependent temperature as
in Eq.(\ref{dD(t)}) above.\\

{\em Steady non-uniform temperature - } Let us now try to
elaborate on the meaning of the above results. For a non-uniform
temperature, Eq.(\ref{dD(r)}) implies that a spherical particle
should undergo anisotropic diffusive motion, with the anisotropy
being inherited from the structure of the temperature profile
\cite{spherical}. For example, a {\em harmonically modulated}
temperature profile $T({\bf r})=T_0+\delta T \cos({\bf k}\cdot{\bf
r})$, yields $\delta D_{ij}({\bf r})=-{k_{\rm B} \delta T k/(64
\eta}) \; (3 \delta_{ij}+{\hat k}_{i} {\hat k}_{j}) \cos({\bf
k}\cdot{\bf r})$ that predicts an anisotropy of $\delta
D_{\parallel}/\delta D_{\perp}=4/3$ for the ratio between the
corrections to the diffusion constant in the parallel and
perpendicular directions respectively. It also shows that for a
small tracer of size $a$ the correction is typically weaker than
the usual term in (1) by a factor of order $\sim (k a) (|\delta
T|/T_0)$, with both terms typically smaller than $1$.

To make connection with experiments, we need to consider realistic
temperature profiles. Non-uniform temperature can be achieved by
locally heating the fluid, and the profile can be obtained through
the heat equation $-\kappa \nabla^2 T({\bf r})=Q({\bf r})$, where
$\kappa$ is the thermal conductivity and $Q({\bf r})$ is the
volume density of heat generated by external sources per unit
time. The heating can originate from Ohmic loss if, for example, a
wire carrying electric current is placed in the fluid, or from
optical heating with an IR laser light focused into a spot within
the fluid. Note, however, that in the latter case we need to make
sure that the dielectric constant of the tracer particle matches
that of the fluid at the wavelength of the beam, or that the
particle is sufficiently far from the spot, so that optical
trapping of the dielectric particle does not interfere with the
effect that we are studying here. We choose a source with the
anisotropic Gaussian profile $Q({\bf r})={P_0/(\ell^2 \ell_z)}
\exp\left\{-\pi{(x^2+y^2)/\ell^2}-\pi {z^2/\ell_z^2}\right\}$, in
which $P_0$ is the total dissipation power, and find a variety of
different asymptotic regimes: (i) For $\ell,\ell_z \ll r$, where
the tracer is sufficiently far from a {\em point-like} heating
source, we find $T({\bf r})=T_0+P_0/(4 \pi \kappa r)$, and
\begin{equation}
\delta D_{ij}({\bf r})=-{1 \over 64 \pi^2}\left({P_0 k_{\rm B} \over \kappa \eta r^2}\right)
\;(2 \delta_{ij}-{\hat r}_i{\hat r}_j),\label{dDijpo}
\end{equation}
to the leading order. Note that $\delta D$ has a different (faster) power law decay with $r$
as compared to that of $T$.
Moreover, we find an anisotropy
$\delta D_{rr}/\delta D_{\perp \perp}=1/2$.
(ii) For $\ell \ll r \ll \ell_z$, where we have a {\em linear} source, we find
$T({\bf r}_{\perp})=T_0-(P_0/2 \pi \kappa \ell_z) \ln|{\bf r}_{\perp}|$,
and corrections to the diffusion tensor as
\begin{eqnarray}
\delta D_{zz}({\bf r}_{\perp})&=&-{3 \over 128 \pi}\left({P_0 k_{\rm B}
\over \kappa \eta \ell_z r_{\perp}}\right), \nonumber \\
\delta D_{ij}^{\perp}({\bf r}_{\perp})&=&-{1 \over 128 \pi}\left({P_0 k_{\rm B}
\over \kappa \eta \ell_z r_{\perp}}\right)
\;(4 \delta_{ij}^{\perp}-{\hat r}_{\perp i}{\hat r}_{\perp j}), \label{dDijli}
\end{eqnarray}
and $\delta D_{z \perp}({\bf r}_{\perp})=0$ to the leading order.
Since in this case the temperature variations in space is very weak
compared to the variations in $\delta D$, it may be a good
candidate for experimental observation of the space dependency of $\delta D$.
We also find an anisotropy
$\delta D_{zz}/\delta D_{\theta \theta}=\delta D_{\rho \rho}/\delta D_{\theta \theta}=3/4$
in terms of the components in cylindrical coordinates.
(iii) For $\ell_z \ll r \ll \ell$, where we have a {\em planar} source, we find
$T(z)=T_0-(P_0/2 \kappa \ell^2) |z|$, and the corrections as
\begin{eqnarray}
\delta D_{zz}(z)&=&{1 \over 16 \pi}\left({P_0 k_{\rm B} \over \kappa \eta \ell^2}\right) \ln|z|,
\nonumber \\
\delta D_{ij}^{\perp}(z)&=& {3 \over 64 \pi}\left({P_0 k_{\rm B} \over \kappa \eta \ell^2}\right) \ln|z|
\;\delta_{ij}^{\perp},\label{dDijpl}
\end{eqnarray}
and $\delta D_{z \perp}(z)=0$ to the leading order.
This case also corresponds to a temperature profile with
a constant gradient, and is expected to have the strongest effects
in terms of magnitude as compared to other
regimes. The anisotropy in this case is found as $\delta D_{zz}/\delta D_{\perp \perp}=4/3$, similar
to the harmonically modulated case.
(iv) Finally, for $\ell,\ell_z \gg r$, where we have an extended anisotropic source with an aspect ratio
$\alpha \equiv \ell_z/\ell$, we find
$T({\bf r})=T_0+(P_0/2 \pi \kappa \ell) \ln\left[\alpha+\sqrt{\alpha^2-1}
\over \alpha-\sqrt{\alpha^2-1}\right]
/\sqrt{\alpha^2-1}$, and
\begin{equation}
\delta D_{ij}({\bf r})=-{1 \over 128 \pi^2} \left({P_0 k_{\rm B}
\over \kappa \eta \ell^2}\right) \Lambda_i
\;\delta_{ij},\label{dDijex}
\end{equation}
to the leading order, where
$\Lambda_z=[(4-3 \alpha^2) \tanh^{-1}(\sqrt{1-\alpha^2})-\sqrt{1-\alpha^2}]/(1-\alpha^2)^{3/2}$,
and
$\Lambda_{\perp}=[(6-7 \alpha^2) \tanh^{-1}(\sqrt{1-\alpha^2})+\sqrt{1-\alpha^2}]
/[2(1-\alpha^2)^{3/2}]$.
This is a very interesting case in the sense that both temperature and diffusion tensor have constant
profiles to the leading order, and yet the diffusing tracer feels
the large-scale anisotropy in the temperature
profile, as manifest in Eq.(\ref{dDijex}). The anisotropy in this case is found as
\begin{equation}
{\delta D_{zz} \over \delta D_{\perp \perp}}=
\frac{2(4-3 \alpha^2) \tanh^{-1}(\sqrt{1-\alpha^2})-2 \sqrt{1-\alpha^2}}
{(6-7 \alpha^2) \tanh^{-1}(\sqrt{1-\alpha^2})+\sqrt{1-\alpha^2}},\label{aspect}
\end{equation}
which depends on the aspect ratio, as plotted in Fig.~1.
Note that Eq.(\ref{aspect}) shows that the anisotropy
asymptotes to $4/3$ in the limit $\ell_z \ll \ell$,
similar to the planar and the harmonically modulated cases.
Moreover, it reveals that oblate profiles lead to stronger anisotropy than the prolate cases.
This trend is in agreement with diminished anisotropy in the case of linear sources
($\delta D_{zz}/\delta D_{\rho \rho}=1$)\\

\begin{figure}[t]
\centerline{\epsfxsize 5cm \rotatebox{-90}{\epsffile{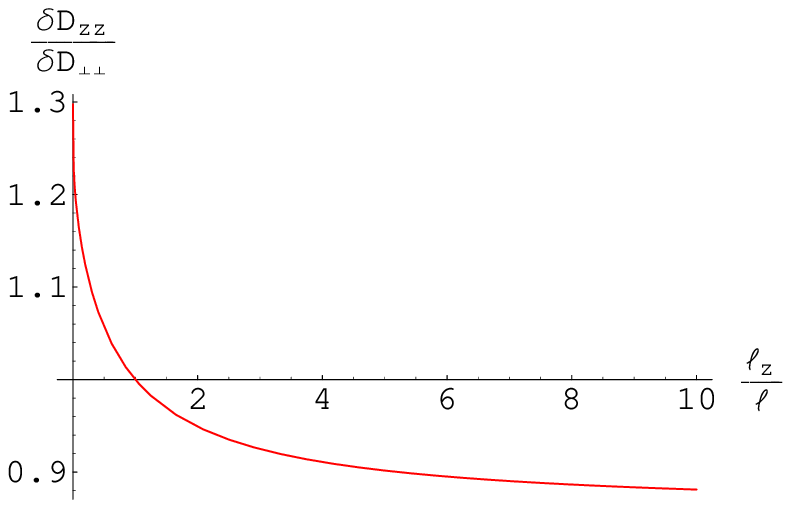}}}
\vskip 1.5truecm
FIG.~1. A plot of the anisotropy in the correction to the diffusion coefficients
Eq.(\ref{aspect}) as a function the
aspect ratio of the heating profile. The anisotropy is more pronounced for oblate profiles
as compared with the prolate cases.
\label{fig1}
\end{figure}

{\em Temporally varying uniform temperature -}
For time dependent temperatures, hydrodynamic retardation leads to non-locality in time,
as evident in Eq.(\ref{dD(t)}) above. This memory effect can be well illustrated in the case of an
oscillating temperature field of the form $T({\bf r})=T_0+\delta T \cos \omega_0 t$, where it leads
to an isotropic correction to the diffusion tensor as $\delta D_{ij}(t)=\delta_{ij} \delta D(t)$ with
\begin{equation}
\delta D(t)={k_{\rm B} \delta T \over 4 \pi \sqrt{2} \;\eta}
\sqrt{\rho \omega_0 \over \eta}\; \cos(\omega_0 t-3 \pi/4).
\end{equation}
The above result shows a systematic phase lag of $3 \pi/4$ for the hydrodynamic correction to the diffusion
constant, with respect to the temperature. The characteristic length in the above equation is set by the
kinematic viscosity $\eta/\rho$
(that has the dimension of a diffusion constant) and the frequency $\omega_0$
as $a_T=\sqrt{\eta/\rho \omega_0}$, and is of the order of $1.3 \;{\rm mm}$ for water at
$\omega_0/(2 \pi)=0.1 \;{\rm Hz}$.\\

{\em Discussion -}
The above results are interesting from a fundamental point of view
because they reveal that hydrodynamic interactions reflect on a
diffusing particle the global knowledge of the temperature
heterogeneity and anisotropy. They may appear though to be
difficult to verify experimentally, because in typical cases the
correction will be considerably smaller than the conventional
value given by Einstein's relation. However, we would like to
point out that due to the anisotropy in the diffusion tensor, one
can choose to probe a quantity such as $D_{xy}$, by direct
measurements of $\langle \Delta x(t) \Delta y(t) \rangle/2 t$, in
a non-symmetrical coordinate system. Since this quantity is always
zero for a spherical tracer in the local picture, it would provide
a feasible experimental test of this effect.

The structure of the stochastic forcing in Eqs.(\ref{NSequ}) and
(\ref{noise}) has a crucial role in determining the form of the
above results. While we are studying systems that are only
slightly away from equilibrium (by choosing a noise correlator
that is a smooth perturbation of the one that ensures thermal
equilibrium), other noise structures could drastically change the
results so that for example the decomposition in Eq.(\ref{Stokes})
could be blurred and the scaling with length could be altered.

Let us recall that our description of the fluid dynamics is
clearly a great simplification, which could be improved by taking
into account the temperature dependence of the density (and thus
release of the incompressibility assumption) and the viscosity,
and also the convection of thermal fluctuations, along the lines
of the approach used in Ref. \cite{Sig} to study fluctuations of a
fluid away from equilibrium. With these ingredients, the
long-range character of the effects described here should then
suffer from screening, but the corresponding screening length will
remain large for weakly compressible fluids.

The hydrodynamic contribution to the effective diffusivity is
independent of the size and shape of the tracer as long as the
shortest length-scale over which the temperature profile varies
significantly is considerably larger than the size of the tracer
particle. In this regard, a proper account of the tracer geometry
for larger particles in strong temperature gradients may become
important, when the interplay with Soret effect is expected to
appear more clearly. We nevertheless expect that some qualitative
features obtained in the present work should remain, i.e. that the
fluctuating character of the particle motion should be sensitive
to the geometry and history of the temperature field.

Eventually, as pointed out in the introduction, the most important
steps to be taken are likely to be on the experimental side with
either direct observation of single tracers, of sets of particles
in a well-controlled fluidic and thermal environment, or direct
probing of the anisotropy in the solvent self-diffusivity (in
which case we expect minimal interference due to Soret effect)
\cite{Por}.

{\em
We are grateful to D. Bartolo and M. Martin for interesting discussions and comments.
One of us (RG) would like to acknowledge ESPCI for hospitality during his visit, and support
through the Joliot visiting chair.}

\end{multicols}

\begin{references}

\bibitem{Mac}
G.S. McNab, A. Meisen, J. Coll. Int. Sci., {\bf 44}, 339 (1973).

\bibitem{Schimpf}
M.E. Schimpf, S.N. Semenov, J. Phys. Chem. B, {\bf 104}, 9935 (2000).

\bibitem{Mic1}
{\em Micro Total Analysis Systems 2000}, Eds.
A. van den Berg, W. Olthuis, P. Bergveld, Kluwer Academic Publishers, Dordrecht (NL), 2000.\\
{\em Micro Total Analysis Systems 2001}, eds. J.M. Ramsey
and A. van den Berg, Kluwer Academic Publishers, Dordrecht (NL), 2001.

\bibitem{Mic2}
A.D. Stroock, G. Whitesides, {\em Physics Today} {\bf 54}, 42-48
(2001).

\bibitem{Ross}
D. Roos, M. Gaitan, L.E. Locascio, p. 239,
in {\em Micro Total Analysis Systems 2001}, ed. J.M. Ramsey
and A. van den Berg, Kluwer Academic Publishers, Dordrecht (NL), 2001.

\bibitem{Pro84}
This is clearly a much simpler model than the one used to study
critical anomalies of thermal diffusivity and kinematic viscosity
in: E. Meron, I. Procaccia, Phys. Rev. A, {\bf 30}, 3221 (1984).

\bibitem{Schmitz}
For a recent review on fluctuations in non-equilibrium steady
states, see: R. Schmitz, Phys. Rep. {\bf 171}, 1 (1988).

\bibitem{regular}
For the regular term that implies a local and instantaneous
application of Einstein's relation in a temperature field with
spatial and temporal heterogeneity, see: E.H. Hauge and A.
Martin-L\"of, J. Stat. Phys. {\bf 7}, 259 (1973).

\bibitem{LanLif}
L.D. Landau and E.M. Lifshitz, {\it Fluid Mechanics} 2nd edition
(Pergamon, Oxford, England, 1987).

\bibitem{Procaccia}
For a discussion on other choices of noise correlation in the
context of turbulence, see: V. L'vov and I. Procaccia, in: {\em
Fluctuating Geometries in Statistical Mechanics and Field Theory},
F. David, P. Ginsparg, and J. Zinn-Justin, eds. (Elsevier,
Amsterdam, 1996).

\bibitem{ramin}
The integral expression is also useful to extract symmetry properties of the mobility
tensor, such as the (factor of 2) anisotropy in the friction coefficient for a slender object. For
related arguments in the analogous case of electrostatic response coefficients such as
capacitance and polarizability, see: R. Golestanian, Phys. Rev. E {\bf 62}, 5242 (2000).

\bibitem{spherical}
The example focuses on the anisotropic contributions in the
universal term caused by global anisotropy in the temperature
field. The specific reference is made to spherical tracers only
because their regular diffusivity is isotropic and thus they will
be good candidates for detecting any possible anisotropy that
originates from the non-local term. This, of course, does not mean
that the results only apply to the case of spherical tracers.

\bibitem{Sig}
A.M. Tremblay, M. Arai, E.D. Siggia, Phys. Rev. A, {\bf 23}, 1451 (1981).

\bibitem{Por}
P. Porion {\em et al.}, Phys. Rev. Lett., {\bf 87}, 208302 (2001).



\end{references}
\end{document}